\documentclass[a4paper,11pt]{article}

\usepackage{fullpage}
\usepackage{physics}
\usepackage{amsmath}
\usepackage{amssymb}
\usepackage{graphicx}
\usepackage{hyperref}
\DeclareGraphicsExtensions{.pdf}

\def\A{\ensuremath{\mathrm{A}}}%
\def\B{\ensuremath{\mathrm{B}}}%
\def\W{\ensuremath{\mathrm{W}}}%
\def\WW{\ensuremath{\mathrm{W^+}\mathrm{W^-}}}%
\def\eL{\ensuremath{e_L}}%
\def\eR{\ensuremath{e_R}}%
\def\e{\ensuremath{e}}%
\def\eegg{\ee\gamma\gamma}%
\def\eeragg{\ee\ra\gamma\gamma}%
\def\ege{\overline{\e_\lambda}\gamma^\mu\e_\lambda}%
\def\ehe{\overline{\eL}\Phi\eR}%
\def\egghe{\overline{\eL}\sigma^{\mu\nu}\Phi\eR}%
\def\Lambdapm{\Lambda_\pm}%

\begin{document}

\title{Precision studies of quantum electrodynamics at future $\epem$ colliders}
\author{J. Alcaraz Maestre \\ CIEMAT-Madrid}
\date{CIEMAT Technical Report 1499
\footnote{ISBN/ISSN: 2695-8864, NIPO: 832-21-019-3, CIEMAT, 2021, \url{https://cpage.mpr.gob.es/}}
} 

\maketitle

\begin{abstract}
We classify the possible deviations from the Standard Model
in the QED-dominated $\eeragg$ process under the assumption of a preserved 
$SU(2)_L \times U(1)_Y$ symmetry. We find that the only deviations really observable
in practice correspond to a correction of the differential cross section by a factor
$(1 + \frac{c_8~s^2}{8\pi\alpha\Lambda^4}~\sin^2\theta)$, where $\Lambda$ is the 
scale of new physics, $\theta$ is the polar angle of any of the final state
photons and $c_8$ is a constant of order 1.
We also provide sensitivity estimates for QED deviations at future
$\ee$ facilities. An $\ee$ collider operating at $\sqrt{s}=3$ TeV could
provide sensitivity to $\Lambda$ scales as large as 15 TeV, provided
that acceptances and efficiencies are controlled at the per mille
level. Finally, we also discuss the possibility of a measurement of the
luminosity at the FCC-ee with $\lesssim 10^{-4}$ precision, using analyses of
the $\ee\ra\gamma\gamma$ process at $\sqrt{s}\approx\MZ$ energies.
\end{abstract}

\section{Introduction}

The successful confrontation of the Standard Model (SM)
\cite{Glashow:1961tr,Weinberg:1967tq,Salam:1968rm} to measurements at present collider
experiments provides strong evidence for an underlying $SU(2)_L \times U(1)_Y$ symmetry in
particle physics interactions. Despite the impressive level of agreement between SM
predictions and experimental observations, there are known weak points in the scheme:
gravitational interactions not considered, unnatural and unexplained origin of mass
hierarchies, insufficient amount of CP violation, implications of the presence of massive
neutrinos, unknown unification scheme of fundamental forces at high energy, \ldots The
search for new physics beyond the SM is one of the main goals of present and future
high-energy experiments and, within this context, classifying and constraining the most
sensitive terms is an important exercise. 

The $\eeragg$ process has been used since long time as a golden channel to test the validity
of QED \cite{Feynman:1948fi,Low:1965ka,Renard:1982ij}. Experimentally, its signature is
clean and systematic uncertainties can be easily controlled
\cite{Heister:2002ut,Abdallah:2004rc,Achard:2001nt,Abbiendi:2002je}. In these latest
analyses, performed at LEP2 energies~\cite{Schael:2013ita}, deviations were quantified in
terms of effective Lagrangians that respect the underlying QED U(1)
symmetry~\cite{Eboli:1991ci}. In this report we present instead a prospect study of the
possible deviations under the assumption of a preserved $SU(2)_L \times U(1)_Y$ symmetry, as
suggested by the overwhelming evidence for the validity of the SM up to the electroweak
scale. As we will see, this scheme leads to a simpler scheme to study and isolate new
physics effects. 
 
The measurement of the $\eeragg$ cross section has also been been suggested as
reference for an ultra-precise determination of the luminosity at future Higgs
factories running at the electroweak scale, owing to the fact that the QCD
uncertainties in this process are expected to contribute at the $\lesssim
10^{-5}$ level~\cite{CarloniCalame:2019dom}. This has to be compared with the
current $\approx 10^{-4}$ relative uncertainties expected from Bhabha
scattering at very low polar angles, which will likely require alignment
precisions of the luminosity monitors at the micron
level~\cite{Dam:2021sdj,Alcaraz:2021zvq}. The advantages and potential issues
related with the use of the $\eeragg$ process as luminometer reference is also
addressed in this report. 

In general, the study of neutral gauge boson pair production offers
information complementary to the $\ee\ra\ffbar$ and $\ee\ra\WW$ processes, and
the observation of new physics effects in $\eeragg$ will either confirm or
rule out some of the possibilities. An interesting example is the possible
existence of strong gravitational interactions at the TeV scale
\cite{ArkaniHamed:1998rs,Agashe:1999qp}, which should manifest simultaneously
in all channels. 

The classification of the relevant operators preserving the $SU(2)_L
\times U(1)_Y$ symmetry is presented in the first sections of the 
report. We discuss which terms are acceptable, redundant or essentially
excluded by present observations. The second part of the study is
devoted to phenomenological and experimental studies assessing the 
new physics reach of the process at future $\ee$ Higgs factories and 
the experimental uncertainties expected in the measurement of the $\eeragg$ 
cross section at a Tera-Z facility like the one proposed at a future 
FCC-ee collider.

\section{$\eegg$ contact interactions and $SU(2)_L \times U(1)_Y$ symmetry \label{sec:dim6}}

We will assume that new physics: a) lies above the electroweak scale, b) 
respects the $SU(2)_L \times U(1)_Y$ symmetry, and c) decouples from 
the SM at low energy. The possible deviations are parametrized
in terms of effective Lagrangians containing inverse
powers of the scale of new physics, $\Lambda$:
\begin{eqnarray}
    {\cal L} & = & {\cal L}_{SM} + 
              \sum_{n=5}^{\infty} \frac{1}{\Lambda^{n-4}} 
                  \left( \sum_{j} c_{nj}~{\cal O}_{nj} \right)
\end{eqnarray}

 The index $n$ is the dimension of the ${\cal O}_{nj}$ operator and the sum on $j$ is
extended to all possible terms respecting the symmetry. Provided that the ${\cal O}_{nj}$
operator is present and is hermitian, the Wilson coefficients $c_{nj}$ are expected be real
and of order~1. In the limit in which new physics lies much above the electroweak scale and
the characteristic scale of the collision, $\Lambda \gg v, \rts$ ($v=246 \GeV$), terms of
dimension $n$ are suppressed at least by factors of order
$(\frac{\max(v,\rts)}{\Lambda})^{n-4}$ and only the Lagrangians with the lowest dimension
become relevant. The following standard notation is employed in the following:

\begin{itemize}
\item left-handed SM iso-doublet containing the electron: $\eL$,
\item right-handed electron iso-singlet: $\eR$,
\item gauge fields associated to $SU(2)_L$: $\W^I_\mu$,
\item field strengths associated to $SU(2)_L$: 
   $\W^I_{\mu\nu} \equiv \partial_\mu \W^I_\nu - \partial_\nu \W^I_\mu +
   g~\epsilon_{IJK} \W^J_\mu \W^K_\nu$, 
\item gauge field associated to $U(1)_Y$: $\B_\mu$,
\item field strength associated to $U(1)_Y$: 
    $\B_{\mu\nu} \equiv \partial_\mu \B_\nu - \partial_\nu \B_\mu$,
\item photon field: 
    $\A_\mu \equiv  \cos\theta_w \B_\mu + \sin\theta_w \W^3_\mu$,
\item $\Zo$ field: 
    $\Zo_\mu \equiv \cos\theta_w \W^3_\mu - \sin\theta_w \B_\mu$,
\item Higgs field: $\Phi$. In the unitary gauge:
  $\Phi \equiv {\displaystyle \frac{1}{\sqrt{2}}} 
     \left( \begin{array}{c} 0 \\ v + H \\\end{array} \right)$,
\item covariant derivatives:
   \begin{itemize}
     \item 
  for fermions and the Higgs particle:
  $D_\mu \equiv \partial_\mu - i \frac{g}{2} \tau^I \W^I_\mu - i g^\prime Y \B_\mu$,
     \item
  for $SU(2)_L$ tensor strengths: 
  $D_\mu \W^I_{\nu\lambda}\equiv \partial_\mu \W^I_{\nu\lambda} + g~\epsilon_{IJK}
      \W^J_\mu \W^K_{\nu\lambda}$,
     \item
  for the $U(1)_Y$ tensor strength:
  $D_\mu \B_{\nu\lambda}\equiv \partial_\mu \B_{\nu\lambda}$,
   \end{itemize}
\item and dual tensors: 
$\tilde{\W}^I_{\mu\nu} \equiv 
         \frac{1}{2}\epsilon_{\mu\nu}\!^{\rho\lambda} \W^I_{\rho\lambda}$,
$\tilde{\B}_{\mu\nu} = \frac{1}{2}\epsilon_{\mu\nu}\!^{\rho\lambda} 
 \B_{\rho\lambda}$.
\end{itemize}. 

   In the previous expressions $g$ and $g^\prime$ are the $SU(2)_L$ and
$U(1)_Y$ coupling constants, respectively, and $\theta_w$ is the
Weinberg angle. The Pauli matrices are denoted by $\tau^I;
I\in\{1,2,3\}$, the value of the hypercharge is $Y$ and the
electromagnetic coupling constant satisfies: $e\equiv 4\pi\sqrt{\alpha}
= g \sin\theta_w$.  We will also assume a massless electron. This assumption 
leads to many simplifications, and it is certainly justified from the
phenomenological and experimental points of view.

\subsection{Effective Lagrangians of dimension six} 

No Lagrangians containing $\eegg$ couplings and satisfying the general conditions described
above can be built at dimension five. At dimension six, it is well known that all potential
$\eegg$ contact terms are redundant with operators of higher order. The reason is that they
can be expressed via classical equations of motion (EOMs) as combinations of other operators
that do not involve fields with two fermions and two
bosons~\cite{Buchmuller:1985jz,Grzadkowski:2010es}. The immediate consequence is that all
new physical effects will be at least of order $\mathcal{O}((energy)^2/\Lambda^4)$. 

Still, one can consider indirect ways of connecting $\ee$ and $\gamma\gamma$ states via
dimension-six operators. The exchange of neutral gauge bosons in the s-channel is excluded
because no anomalous triple neutral gauge boson couplings of dimension six preserving
$SU(2)_L \times U(1)_Y$ exist. The exchange of Higgs particles is another possibility, but a
large anomalous coupling to electrons is in contradiction with the absence of observed
effects in the $\ee\ra\ee$ cross section for Higgs masses up to the weak scale. Finally, one
expects contributions from operators of the electric or magnetic dipole type: $\sim (\eL
\sigma^{\mu\nu}\Phi \eR) F_{\mu\nu}$. They provide effects of order
$\mathcal{O}(v^2s/\Lambda^4)$ because they connect electron states with different chirality
and therefore do not interfere with the SM amplitude. They are suppressed with respect to
other physical effects at a similar scale because they must be created via loop level
diagrams~\cite{Arzt:1994gp}, and present a dependence as a function of the polar angle
equivalent to that of the dimension-eight operators that are discussed in the next 
section. Last but not least, they are extremely constrained by low-energy precision
experiments~\cite{ACME:2018yjb,Hanneke:2008tm}. 

\subsection{Effective Lagrangians of dimension eight} 

No Lagrangians of dimension seven with $\eegg$ couplings and satisfying the condition of
$SU(2)_L \times U(1)_Y$ invariance can be built~\cite{Lehman:2014jma}. At dimension eight,
and despite the apparent increase in the number of possibilities, most of the potential
terms can be effectively ignored, thus easing the classification task. CP-odd terms and all
operators that connect electrons of different chirality via $\ehe$, $\egghe$ components can
be dropped. They give contributions of order $\mathcal{O}((energy)^4/\Lambda^8)$ because they
do not interfere with the SM process. The application of EOMs also helps in reducing the
number of possible structures. 

The starting point is a fermionic component of the type $\ege$. It can be complemented with
covariant derivatives ($D^\mu$), gauge fields ($F^{\mu\nu},\tilde{F}^{\mu\nu}$) and Higgs
fields by pairs ($\Phi^\dagger,\Phi$), but it only terms with one covariant derivative and
two gauge fields lead to genuine $\eegg$ coupling terms. Terms with five or three
derivatives are equivalent to terms with more gauge fields and less derivatives. Using
partial integration, the EOMs and the Bianchi identities we can also ignore in the 
classification terms containing derivatives of $F^{\mu\nu}$ and $\tilde{F}^{\mu\nu}$ fields.
The following final set is obtained:

\begin{eqnarray}
  {\cal O}_{\eR\eR\B\B} & = &
     (i~\bar{\eR} \gamma_\mu D_\nu  \eR) \B^{\mu\rho}~\B_\rho^{~\nu}+~{h.~c.} 
                    \label{eq:dim8_1} \\
  {\cal O}_{\eL\eL\B\B} & = &
     (i~\bar{\eL} \gamma_\mu D_\nu  \eL) \B^{\mu\rho}~\B_\rho^{~\nu}+~{h.~c.} 
                    \\
  {\cal O}_{\eR\eR\W\W} & = &
(i~\bar{\eR} \gamma_\mu D_\nu  \eR) 
              \W^{I~\mu\rho}~\W_{~\rho}^{I~\nu}+~{h.~c.} 
                    \\
  {\cal O}_{\eL\eL\W\W} & = &
(i~\bar{\eL} \gamma_\mu D_\nu  \eL) 
              \W^{I~\mu\rho}~\W_{~\rho}^{I~\nu}+~{h.~c.} 
                    \\
  {\cal O}_{\eL\eL\W\B} & = &
(i~\bar{\eL} \tau^I \gamma_\mu D_\nu  \eL) 
              \W^{I~\mu\rho}~\B_\rho^{~\nu}+~{h.~c.} 
                    \\
  {\cal O}_{\eL\eL\B\W} & = &
(i~\bar{\eL} \tau^I \gamma_\mu D_\nu  \eL) 
              \B^{\mu\rho}~\W_{~\rho}^{I~\nu}+~{h.~c.} 
                    \label{eq:dim8_6}
\end{eqnarray}

All operators \ref{eq:dim8_1}-\ref{eq:dim8_6} are structurally similar.
Consequently, only the following structure can contribute to the
$\eeragg$ process at order $\mathcal{O}((energy)^2/\Lambda^4)$ and at the same 
time respect the underlying SM symmetry:

\begin{eqnarray}
  {\cal O}_{\e\e\gamma\gamma} & \rightarrow &
 \left[ i~\bar{\e} \gamma_\mu
   \frac{1 \pm \gamma_5}{2} \partial_\nu \e \right]
         \A^{\mu\rho}~\A_\rho^{~\nu}+~{h.~c.} 
\end{eqnarray}

\subsection{Beyond dimension eight} 

The next level of deviations from the SM corresponds to operators of
dimension ten. We may create new $\eegg$ couplings by combining
operators of dimension eight with additional Higgs fields or covariant
derivatives by pairs.  The leading corrections in this case are of
order $\mathcal{O}((energy)^3/\Lambda^6)$, which can be absorbed via a
redefinition of the dimension-eight couplings via form factors. 

Beyond dimension ten, one should consider not only dimension-twelve Lagrangians, but also
the dimension-eight terms with contributions of order $\mathcal{O}((energy)^4/\Lambda^8)$.
Although highly suppressed, some of these operators ( $\overline{\e_L}\Phi\e_R
\A^{\mu\nu}\A_{\mu\nu}$, for instance), provide differential cross sections that are
experimentally distinguishable from the one predicted by the ${\cal O}_{\e\e\gamma\gamma}$
Lagrangian. 

\section{A simple behavior for the leading deviations in the $\eeragg$
process} 

According to the previous discussion, the lowest-order deviations in
the SM $\eeragg$ process allowed by $SU(2)_L \times U(1)_Y$ invariance
can be quantified using the following Lagrangian:

\begin{eqnarray}
  {\cal O}_8 & = & 
     \frac{1}{\Lambda^4} \left[ \bar{\e} \gamma_\mu 
     \left(f_{8L} \frac{1 -\gamma^5}{2} + 
          f_{8R} \frac{1 +\gamma^5}{2}\right)
           i\partial_\nu \e \right]~
         \A^{\mu\rho}~\A_\rho^{~\nu}+~{h.~c.}
\end{eqnarray}

\noindent where $f_{8L},f_{8R}$ are constants of order unity and
$\Lambda$ is the possible scale of new physics. The resulting
differential cross section, including SM and new physics contributions,
is:

\begin{eqnarray}
  \frac{d\sigma}{d\Omega} & = & 
          \frac{1}{2} \left(\frac{d\sigma}{d\Omega}\right)_{SM}
         \left[
\left(1 + \frac{f_{8L} s^2}{16\pi\alpha\Lambda^4} \sin^2\theta\right)^2
                               + 
\left(1 + \frac{f_{8R} s^2}{16\pi\alpha\Lambda^4} \sin^2\theta\right)^2
                        \right]
\end{eqnarray}

\noindent where $\theta$ is the polar angle between the photon and any
of the beam particles, and $\Omega$ the solid angle, with $\cos\theta$
defined in the $[0,1)$ range. The term $(d\sigma/d\Omega)_{SM}$ is  
the Born-level QED differential cross section:

\begin{eqnarray}
{\displaystyle \left(\frac{d\sigma}{d\Omega}\right)_{SM}} & = &
{\displaystyle \frac{\alpha^2}{s}~\frac{1+\cos^2\theta}{\sin^2\theta}}
\end{eqnarray}

As expected, the effect is similar to equivalent dimension-8 $\eegg$ contact terms that were
previously proposed~\cite{Boudjema:1989gc,Mery:1987et,Mery:1986qn} and, at first order of
deviation ($\propto s^2/\Lambda^4$), to the one predicted by the inclusion of QED cutoff
parameters $\Lambdapm$ in the electron propagator \cite{Feynman:1948fi}. Indeed, the
$\Lambda$ scale coincides numerically with $\Lambdapm$ for new interactions with vector-like
couplings of electromagnetic size: 

\begin{eqnarray}
      c_8 = f_{8l} = f_{8R} = \pm e^2 & \Rightarrow &  ~~~~
  \frac{d\sigma}{d\Omega} = 
          \left(\frac{d\sigma}{d\Omega}\right)_{SM}
\left(1 \pm \frac{s^2}{2\Lambdapm^4} \sin^2\theta + 
\frac{s^4}{16\Lambdapm^8} \sin^4\theta \right)
\end{eqnarray}

Note that the leading $\mathcal{O}(s^2/\Lambdapm^4)$ deviations are
either positive or negative depending on the sign of $c_8$, but the
total cross section always stays positive. 

Reducing the study of all dominant QED deviation effects to the study
of only one operator ($\mathcal{O}_8$) has strong implications. First,
any conceivable effect on the $\eeragg$ process from decoupled new
physics must be of order $\mathcal{O}(s^2/\Lambda^4)$ or smaller.
Second, the only allowed deviation at this order is a relative increase
or decrease of the differential cross section in the central region of
detectors, precisely following a $\sin^2\theta$ behaviour. Examples of
new physics searches in this channel are excited electrons
\cite{Low:1965ka,Renard:1982ij,Hagiwara:1985wt} or the possible
manifestation of extra-dimensional gravity effects at the TeV scale
\cite{ArkaniHamed:1998rs,Agashe:1999qp,Giudice:1998ck}. Both types of
signals indeed follow the required dependence at order
$\mathcal{O}(s^2/\Lambda^4)$, despite their totally different physics
origin (magnetic-like coupling in t-channel versus spin-2 exchange is
s-channel).
   
Some of the new Lagrangians experimentally exploited at
LEP~\cite{Schael:2013ita}, denoted by $\Lambda_6,\Lambda_7$ and
$\Lambda_8$, were proposed in Reference~\cite{Eboli:1991ci}. By
construction they were not expected to respect the $SU(2)_L \times
U(1)_Y$ symmetry, but just the U(1) QED symmetry. However, it is 
illustrative to understand why they do not appear to be relevant in 
our classification:

\begin{itemize}

\item {${i (\bar{\e} \gamma_\mu \overleftrightarrow{D}_\nu  \e) (g_6 F^{\mu\nu} + \tilde{g}_6 \tilde{F}^{\mu\nu})}$:} 
this Lagrangian is redundant at order $\mathcal{O}(s/\Lambda^2)$ 
simply because it has dimension 
six~\cite{Grzadkowski:2010es}. This is
confirmed by the leading $\Lambda^{-4}$ dependence of the deviations 
and its angular dependence, which is exactly reproduced by the 
dimension-eight operator ${\cal O}_8$ introduced above.

\item {${\frac{1}{4} \overline{\e}(g_7^S F^{\mu\nu} + i g_7^P\gamma_5
\tilde{F}^{\mu\nu}) \e F_{\mu\nu}}$:} this dimension-seven Lagrangian
is not $SU(2)_L$-invariant, unless we add a Higgs field connecting the
fermionic fields. After that modification it becomes a Lagrangian of
dimension eight. The predicted deviations are suppressed
by an additional factor $\frac{v^2}{2\Lambda^2}$, due not only to the 
inclusion of the Higgs field, but also to the lack of interference with 
the Standard Model process (it connects electron states of different 
chirality). The combined effect is an operator with a leading contribution 
of order $\mathcal{O}(s^4/\Lambda^8)$, which was neglected in our 
classification of leading deviation terms.

\item {${\frac{1}{8} \overline{\e}\gamma^\mu(g_8^V - g_8^A\gamma_5 \e)
(\partial_\mu \tilde{F}^{\alpha\beta}) F_{\alpha\beta}}$;} this
dimension-eight Lagrangian is insensitive to new physics
from a practical point of view. It is related via the classical 
equations of motion with a Lagrangian proportional to the electron mass. 
This explains the almost negligible contribution obtained in 
Reference~\cite{Eboli:1991ci}, the poor scale limits obtained at LEP and 
why it was discarded a priori in our approach.

\end{itemize}

\section{Expectations at future $\ee$ colliders}

We will first evaluate the statistical sensitivity at different
energies and luminosities assuming an extended likelihood fit to the
differential distribution: 

\begin{eqnarray}
      \frac{d\sigma}{d\cos\theta} & = & \frac{2\pi\alpha^2}{s}~\frac{1+\cos^2\theta}{\sin^2\theta}~\left[1 + \lambda~\frac{s^2}{2}\sin^2\theta\right] \label{eq:diff-lik}
\end{eqnarray}

\noindent where $\cos\theta$ is defined in the $[0,1)$ range and the   
parameter $\lambda$ provides a direct connection with the
electromagnetic cut-off parameter limits obtained in past experiments:
$\lambda\equiv \pm 1/\Lambdapm^4 \equiv \pm f/(e^2\Lambda^4)$. 

In a real high-precision analysis, one will take into the exact shape of the differential
distribution including higher order corrections (see also discussion in the next section)
and adopt an appropriate definition of the event polar angle to account for cases where
additional hard photons are emitted. At LEP~\cite{Schael:2013ita} it was found that a simple
definition that implicitly assumes additional beam-collinear radiation was adequate:
$\cos\theta \equiv \tanh(|y_1-y_2|/2)$, where $y_1$ and $y_2$ are the rapidities of the two
high-energy selected photons. This should probably be complemented in the future with an
additional boost of the system in the transverse direction to account for cases with
additional non-collinear photons, similarly to the Collins-Soper treatment typically used at
hadron colliders~\cite{CollinsSoper}. 

LEP studies also showed that an acollinearity cut between the two selected photons provided 
a better agreement between the true distribution and a pure Born-level treatment,
mostly at high polar angles~\cite{Schael:2013ita}. These details should be of course 
polished in the future, using more precise theoretical calculations - beyond the
permille level of accuracy - once they become available. In our study, and given
the fact that we are only interested in the level of precision that can be
be reached, a pure Born-level description will be assumed. A likelihood fit to the previous
angular distribution of Equation~\ref{eq:diff-lik} in the absence of signal ($\lambda=0$) 
leads to the following uncertainty on $\lambda$: 

\begin{eqnarray}
      \Delta\lambda & = & \frac{2}{s^2\sqrt{<\sin^4\theta>}}~\frac{1}{\sqrt{N_{ev}}}
\end{eqnarray}

\noindent where $N_{ev}$ is the number of selected $\gamma\gamma$ events and $<>$ denotes an
average value over this sample. The statistical uncertainty $\Delta\lambda$ was determined
assuming a parabolic behavior around the minimum of the fit. This is justified because
$\Delta\lambda$ is expected to be rather small for the integrated luminosities of future
colliders. Assuming a large constant acceptance ($\approx 100\%$) and a measurement in the
region $\cos\theta<c_0$, an approximate estimate of $N_{ev}$ and $<\sin^4\theta>$ is given
by: 

\begin{eqnarray}
      N_{ev} & \approx & L~\frac{2\pi\alpha^2}{s}~\left( \log(\frac{1+c_0}{1-c_0})- c_0 \right) \; ,\\
      <\sin^4\theta> & \approx & 
      \frac{\left(c_0 - \frac{c_0^5}{5}\right)}{\left(\log(\frac{1+c_0}{1-c_0})- c_0\right)}
\end{eqnarray}

\noindent where $L$ is the integrated luminosity. The results of the
fit for different $\ee$ colliders, energies and proposed luminosities
is collected in Table~\ref{tab:Lambdapm}, for a cut $c_0=0.95$.

\begin{table}[htb]
\begin{center}

\begin{tabular}{|c|c|c|c|c|c|c|} \hline
Collider & $\sqrt{s}$ & $L$ & $\Delta\lambda$ & $\Delta\sigma_{NP}/\sigma_{SM}$ & $\Lambdapm$ limit & $\Lambda$ limit\\
option & [TeV] & [ab$^{-1}$] & [TeV$^{-4}$] & & [TeV] & [TeV] \\
\hline\hline
LEP & $<0.21$ & $0.003$ & $^{+24}_{-23}$ & - & $0.431/0.339$ & $0.783/0.616$ \\
\hline
FCC-ee & $0.09$ & $          150.0$ & $6.7\times 10^{-1}$ & $1.1\times 10^{-5}$ & $0.9$ & $\phantom{1}1.7$ \\
FCC-ee & $0.16$ & $\phantom{1}10.0$ & $4.8\times 10^{-1}$ & $7.2\times 10^{-5}$ & $1.1$ & $\phantom{1}1.8$ \\
FCC-ee & $0.24$ & $\phantom{11}5.0$ & $2.0\times 10^{-1}$ & $1.5\times 10^{-4}$ & $1.3$ & $\phantom{1}2.3$ \\
FCC-ee & $0.35$ & $\phantom{11}1.5$ & $1.2\times 10^{-1}$ & $4.0\times 10^{-4}$ & $1.4$ & $\phantom{1}2.6$ \\
CEPC & $0.09$ & $\phantom{1}16.0$ & $2.0\times 10^{-1}$ & $3.2\times 10^{-5}$ & $0.7$ & $\phantom{1}1.3$ \\
CEPC & $0.16$ & $\phantom{11}2.6$ & $9.4\times 10^{-1}$ & $1.4\times 10^{-4}$ & $0.9$ & $\phantom{1}1.6$ \\
CEPC & $0.24$ & $\phantom{11}5.6$ & $1.9\times 10^{-1}$ & $1.4\times 10^{-4}$ & $1.3$ & $\phantom{1}2.3$ \\
ILC    & $0.25$ & $\phantom{11}2.0$ & $2.8\times 10^{-1}$ & $2.5\times 10^{-4}$ & $1.2$ & $\phantom{1}2.1$ \\
ILC    & $0.50$ & $\phantom{11}4.0$ & $2.5\times 10^{-2}$ & $3.5\times 10^{-4}$ & $2.1$ & $\phantom{1}3.8$ \\
CLIC   & $0.38$ & $\phantom{11}1.0$ & $1.1\times 10^{-1}$ & $5.4\times 10^{-4}$ & $1.4$ & $\phantom{1}2.6$ \\
CLIC   & $1.50$ & $\phantom{11}1.5$ & $1.5\times 10^{-3}$ & $1.7\times 10^{-3}$ & $4.3$ & $\phantom{1}7.8$ \\
CLIC   & $3.00$ & $\phantom{11}5.0$ & $1.0\times 10^{-4}$ & $1.9\times 10^{-3}$ & $8.3$ & $15.2$ \\
\hline
\end{tabular}

\caption{Achievable statistical uncertainties on $\lambda$, 68\% CL
relative cross section variations due to potential new physics effects,
$\Delta\sigma_{NP}/\sigma_{SM}$, and 95\% confidence level on the
scales $\Lambdapm,\Lambda$ in the absence of QED deviations. The values
are determined for different $\ee$ collider options, energies and
luminosities. We assume an angular fiducial cut of $\cos\theta<0.95$
and an acceptance approaching 100\%. The present combined LEP
results~{\protect \cite{Schael:2013ita}} are also shown for comparison.}

\label{tab:Lambdapm}

\end{center}
\end{table}

As expected, the sensitivity to new physics increases dramatically with
the collision energy, due to the $s^2/\Lambdapm^4$ dependence. 
A CLIC collider at $\sqrt{s}=3$~TeV will be
sensitive to scales as large as 15 TeV if acceptances and efficiencies
can be controlled at the per mille level
($\Delta\sigma_{NP}/\sigma_{SM}$ column in Table~\ref{tab:Lambdapm}). At
collision energies below 1 TeV systematic effects will have to be
controlled at the $10^{-4}$ level or so in order to fully profit from
the available statistics. For FCC-ee measurements at the Z pole and at
the WW thresholds, the luminosity uncertainty, $\Delta L \approx
10^{-4}$, is therefore a limiting factor.

One could also search for new physics effects using exclusively the shape 
of the differential distribution, in order to be independent of luminosity 
uncertainties. The likelihood function is in this case not extended, and the 
corresponding expression for $\Delta\lambda$ is:

\begin{eqnarray}
      \Delta\lambda & = & \frac{2}{s^2\sqrt{<\sin^4\theta>-<\sin^2\theta>^2}}~\frac{1}{\sqrt{N_{ev}}}
\end{eqnarray}

\noindent with:

\begin{eqnarray}
      <\sin^2\theta> & \approx & 
      \frac{\left(c_0 + \frac{c_0^3}{3}\right)}{\left(\log(\frac{1+c_0}{1-c_0})- c_0\right)}
\end{eqnarray}

The results of this alternative fit are reported in
Table~\ref{tab:Lambdapm2}. While the sensitivity to new physics effects
is reduced by almost a factor of two as expressed in terms of the $\lambda$ parameter, 
there is just a mild $\approx 15\%$ degradation in the $\Lambdapm$ limit.

\begin{table}[htb]
\begin{center}

\begin{tabular}{|c|c|c|c|c|c|c|} \hline
Collider & $\sqrt{s}$ & $L$ & $\Delta\lambda$ & $\Delta\sigma_{NP}/\sigma_{SM}$ & $\Lambdapm$ limit & $\Lambda$ limit \\
option & [TeV] & [ab$^{-1}$] & [TeV$^{-4}$] & & [TeV] & [TeV] \\
\hline\hline
FCC-ee & $0.09$ & $          150.0$ & $12.4\times 10^{-1}$ & $1.9\times 10^{-5}$ & $0.8$ & $\phantom{1}1.4$ \\
FCC-ee & $0.16$ & $\phantom{1}10.0$ & $\phantom{1}8.9\times 10^{-1}$ & $1.3\times 10^{-4}$ & $0.9$ & $\phantom{1}1.6$ \\
FCC-ee & $0.24$ & $\phantom{11}5.0$ & $\phantom{1}3.7\times 10^{-1}$ & $2.8\times 10^{-4}$ & $1.1$ & $\phantom{1}2.0$ \\
FCC-ee & $0.35$ & $\phantom{11}1.5$ & $\phantom{1}2.2\times 10^{-1}$ & $7.5\times 10^{-4}$ & $1.2$ & $\phantom{1}2.2$ \\
CEPC & $0.09$ & $\phantom{1}16.0$ & $37.9\times 10^{-1}$ & $5.9\times 10^{-5}$ & $0.6$ & $\phantom{1}1.1$ \\
CEPC & $0.16$ & $\phantom{11}2.6$ & $17.4\times 10^{-1}$ & $2.6\times 10^{-4}$ & $0.7$ & $\phantom{1}1.3$ \\
CEPC & $0.24$ & $\phantom{11}5.6$ & $\phantom{1}3.5\times 10^{-1}$ & $2.7\times 10^{-4}$ & $1.1$ & $\phantom{1}2.0$ \\
ILC    & $0.25$ & $\phantom{11}2.0$ & $\phantom{1}5.2\times 10^{-1}$ & $4.6\times 10^{-4}$ & $1.0$ & $\phantom{1}1.8$ \\
ILC    & $0.50$ & $\phantom{11}4.0$ & $\phantom{1}4.6\times 10^{-2}$ & $6.5\times 10^{-4}$ & $1.8$ & $\phantom{1}3.3$ \\
CLIC   & $0.38$ & $\phantom{11}1.0$ & $\phantom{1}2.1\times 10^{-1}$ & $9.9\times 10^{-4}$ & $1.2$ & $\phantom{1}2.3$ \\
CLIC   & $1.50$ & $\phantom{11}1.5$ & $\phantom{1}2.8\times 10^{-3}$ & $3.3\times 10^{-3}$ & $3.7$ & $\phantom{1}6.7$ \\
CLIC   & $3.00$ & $\phantom{11}5.0$ & $\phantom{1}1.9\times 10^{-4}$ & $3.5\times 10^{-3}$ & $7.2$ & $13.0$ \\
\hline
\end{tabular}

\caption{Estimated statistical uncertainties and limits from an
alternative non-extended likelihood fit. The fit is a bit less
sensitive than the extended likelihood fit used in
Table~\ref{tab:Lambdapm} but, since only shape effects are taken into
account, the results are insensitive to luminosity uncertainties.}

\label{tab:Lambdapm2}

\end{center}
\end{table}

\section{Measuring the luminosity using $\eeragg$ events}

The $\ee\ra\gamma\gamma$ process has been proposed as alternative reference reaction for
a precise determination of the luminosity at the FCC-ee, given its minimal
dependence on hadronic corrections ($\approx 10^{-5}$~\cite{CarloniCalame:2019dom}).
Compared with standard measurements of luminosity based on the counting of Bhabha events at
low momentum transfer, the $\eeragg$ reaction can be measured at relatively high polar 
angles, where the required precision on the detector acceptance is less 
stringent~\cite{Dam:2021sdj,Alcaraz:2021zvq}. Using a 
simple Born-level estimate and assuming a total integrated luminosity of $150~\text{ab}^{-1}$ at the Z pole we obtain a statistical precision of 
$1\sqrt{N}= 1.3\times 10^{-5}$ for $|\cos\theta|<0.95$, i.e. in a region fully covered by 
the tracker system, and $1\sqrt{N}= 2.0\times 10^{-5}$ for $|\cos\theta|<0.7$, a 
typical region spanned by both barrel tracker and barrel electromagnetic calorimeters. 
This implies a statistical precision well below $10^{-4}$ per year of running at the Z 
pole is a realistic target.

Nevertheless, a luminosity measurement using this channel will be a sensible option only if 
``new physics'' effects can be excluded or at least circumvented.
At the Z pole, and assuming that scales $\Lambda\lesssim 0.7$~TeV are already excluded 
by LEP2, an analysis similar to the one employed to obtain Table~\ref{tab:Lambdapm2} 
would provide uncertainties due to unknown new physics effects never larger than
$\Delta\sigma/\sigma_{SM} \approx 4\times 10^{-4}$. This implies that luminosity 
uncertainties below the permille level are a very realistic target. A way
to further reduce this ``new physics'' uncertainty would be to 
perform a simultaneous fit to both the integrated luminosity and the size of QED deviations, 
following the strategy of decoupling normalization and shape-only effects 
described in the previous section. According to the results of Table~\ref{tab:Lambdapm2}, 
this uncertainty can be reduced at the $\Delta\sigma/\sigma_{SM} \lesssim 2.0\times 10^{-5}$
level for the full sample, and certainly below the $10^{-4}$ level for yearly running 
periods. This decoupling strategy would also be more consistent with current proposals 
that ask for an integrated treatment of NLO corrections and theoretical 
uncertainties within an SMEFT context~\cite{Passarino:2016pzb}. For instance, we note that 
the SM NLO weak corrections for this channel have the same angular behaviour as 
genuine QED deviations, and should therefore be disentangled carefully. 
These NLO effects contribute at the permille level at the Z peak, and at the percent 
level at WW, HZ or $\ttbar$ thresholds~\cite{CarloniCalame:2019dom}. 

One can also investigate whether equivalent or 
better limits have already been obtained with dedicated studies at
the LHC. Elastic proton-proton scattering (interpreted as a $\gamma^*\gamma^*\ra\epem$ 
reaction) is one of the possibilities.
Current results are based on rather limited statistics and have a
non-negligible systematic contribution from single proton dissociation
events~\cite{Cms:2018het}. The expectations from inelastic
collisions for $\ee\gamma\gamma$ contact terms via photonic or leptonic
PDFs seem equally limited~\cite{Manohar:2017eqh,Buonocore:2020nai}.
If we assume fermion universality for the $\ffbar\gamma\gamma$ contact terms 
the situation improves. For instance, a reinterpretation of the
GRW~\cite{Giudice:1998ck} CMS search for large
extra-dimensions~\cite{Sirunyan:2018wnk} provides a direct limit on
$\Lambda$. CMS sets a limit on the extra-dimension GRW scale of $M_S >
7.8$~TeV, which translates into a limit of $\Lambda =  M_S/\sqrt{2} >
5.5$~TeV, well above the expected FCC-ee reach.

Experimental uncertainties and backgrounds must be understood to a similar level of
precision. An almost background-free analysis can be performed by selecting events with zero
tracking activity whatsoever around the two most energetic selected photons. The small 
remaining background from Bhabha events (despite their large cross section) can
be quantified using a control Bhabha sample
where one of the final electrons is selected with extremely tight criteria and the second
electron is misidentified as a photon, due to either ``internal'' bremsstrahlung or
real bremsstrahlung in the detector. Diphoton events lost due to photon
conversions on both sides of the event can be estimated using a control sample of events 
where the conversion affects only one of the two photons, as previously done at LEP. While 
this control sample and the previous Bhabha control
sample present an overlap, they can easily be disentangled by analyzing the energy 
spectrum of the charged electron (smaller on average in the case of conversions). 

Last but not least, a precise knowledge of the detector acceptance requires a precise 
position of the edges of the measurement, at the level of $10-100~\mu m$ over distances of 
a meter, depending on the luminosity precision target ($10^{-5}-10^{-4}$). 
Those precisions can only be obtained by a continuous monitoring of the 
electromagnetic calorimeter positions using charged tracks from Bhabha events.
Note that such a procedure also monitors the size and density of collisions 
in the beam interaction region, which also affects the acceptance. 
All these experimental details should be investigated in more detail 
using realistic full simulations of the FCC-ee detectors, in order to 
finally establish whether a luminosity measurement at the $10^{-4}$ or better 
is feasible or not.

\section{Conclusions}

Under the assumption of a preserved $SU(2)_L \times U(1)_Y$ symmetry,
and in the massless electron limit, any possible deviation from the
Standard Model in the $\eeragg$ process is suppressed at least by
factors of order $\mathcal{O}(s^2/\Lambda^4)$, where $\Lambda$
represents the characteristic scale of new physics. The differential
cross section as a function of the photon polar angle at this order
must follow the dependence:

\begin{eqnarray*}
\left( \frac{d\sigma}{d\cos\theta} \right)_{SM+new} =
\left( \frac{d\sigma}{d\cos\theta} \right)_{SM}~
  \left[ 1 + \frac{c_8~s^2}{8\pi\alpha\Lambda^4}~\sin^2\theta \right]
\end{eqnarray*}

\noindent where $c_8$ is a constant of order 1. We find that any
different behaviour of the differential cross section will be hardly
observable, since it is suppressed by additional $({\rm
energy}/\Lambda)^4$ factors.

We have provided sensitivity estimates for QED deviations at future
$\ee$ facilities. A CLIC collider operating at $\sqrt{s}=3$ TeV could
provide sensitivity to $\Lambda$ scales as large as 15 TeV, provided
that acceptances and efficiencies are controlled at the per mille
level. Finally, we have discussed in some detail the interplay between
potential new physics effects and a possible measurement of the
luminosity at the FCC-ee with $<10^{-4}$ precision, using analyses of
the $\ee\ra\gamma\gamma$ process at the Z pole.


\bibliographystyle{myutphys}
\bibliography{GG_ciemat}{}

\end{document}